# All-optical switch and logic gates based on all-dielectric hybrid silicon-Ge$_2$Sb$_2$Te$_5$ metamaterials

**Zhaojian Zhang,**[1] **Junbo Yang,**[2,*] **Wei Bai,**[3] **Yunxin Han,**[2] **Xin He,**[2] **Jie Huang,**[1] **Dingbo Chen,**[1] **Siyu Xu,**[1] **and Wanlin Xie**[1]

[1] *College of Liberal Arts and Sciences, National University of Defense Technology, Changsha 410073, China*
[2] *Center of Material Science, National University of Defense Technology, Changsha 410073, China*
[3] *Institute of Optics and Electronics Chinese Academy of Sciences, Chengdu 610209, China*
*\* yangjunbo@nudt.edu.cn*

**Abstract:** We numerically propose an all-dielectric hybrid metamaterial (MM) to realize all-optical switch and logic gates in shortwave infrared (SWIR) band. Such MM consists of one silicon rod and one Ge$_2$Sb$_2$Te$_5$ (GST) rod pair. Utilizing the transition from amorphous to crystalline state of GST, such MM can produce electromagnetically induced transparency (EIT) analogue with active control. Based on this, we realized all-optical switching at 1500 nm with a modulation depth 84%. Besides, three different logic gates, NOT, NOR and OR, can also be achieved in this device simultaneously. Thanks to the reversible and fast phase transition process of GST, this device possesses reconfigurable ability as well as fast response time, and has potential applications in future optical networks.

## 1. Introduction

During the past decades, metamaterials (MMs) have drawn much attention due to the ability to manipulate light in subwavelength scale [1]. The original meta-resonators of MMs are usually made of noble metals such as silver and gold, utilizing surface plasmon polaritons (SPPs) to construct resonances on the surfaces [2]. However, such plasmonic MMs suffer from ohmic loss, induced heating and CMOS incompatibility [3]. Recently, all-dielectric MMs, which is based on Mie resonances in the dielectric meta-resonators with high refractive indices, have emerged as potential roles to compensate for those weaknesses. Dielectric nanostructures can support both electric and magnetic resonances with low intrinsic loss, and are more cost-effective [4]. These advantages make all-dielectric MMs become promising CMOS-compatible platforms to realize various photonic applications such as enhanced high-harmonic generation [5-7], lightwave manipulation [8-10], high-sensitive sensing [11-13] and topological insulators [14-16].

In practice, active devices are important since they can provide on-demand and tunable functionalities. For plasmonic MMs to be active, one of most common strategies is to hybridize metal meta-resonators with active optical materials, such as semiconductors [17-19], two-dimensional (2D) materials [20-22] and phase-change materials (PCMs) [23-25]. Those so-called hybrid MMs can bring novel functions with active control, and have become a rising branch of MMs. Similarly, all-dielectric active hybrid MMs with low loss also need to be explored [26]. Up to now, liquid crystal [27,28] and PCM Ge$_2$Sb$_2$Te$_5$ (GST) [29-31] have been integrated as efficient active dielectrics. However, most related works utilize active media as claddings, substrates or meta-elements, few works have attempted to adopt hybrid dielectric meta-resonators.

In this work, all-dielectric hybrid metamaterials based on GST and silicon are studied numerically. Since GST and silicon have close refractive indices in shortwave infrared (SWIR) band, Mie resonances with same resonant wavelengths can exist in GST and silicon nanoparticles with close footprints. Therefore, we propose hybrid dielectric meta-resonators in a periodic unit cell, including one silicon rod and two GST rods, to realize actively tunable electromagnetically induced transparency (EIT) analogue. Utilizing the phase transition of GST, i.e., from amorphous state (aGST) to crystalline state (cGST), the interference between the electric dipole mode in silicon rod and the electric quadrupole in GST rods can be influenced, enabling the switch and logic gate functions. Benefitting from the optical controllability, reversibility and ultrafast transition time (in nanosecond or even sub-nanosecond) of phase transition process [32], such metadevice is reconfigurable, meanwhile possesses ultrafast response time. This work can find potential applications in future nanophotonic systems for optical communication and information processing.

## 2. Structures, materials and methods

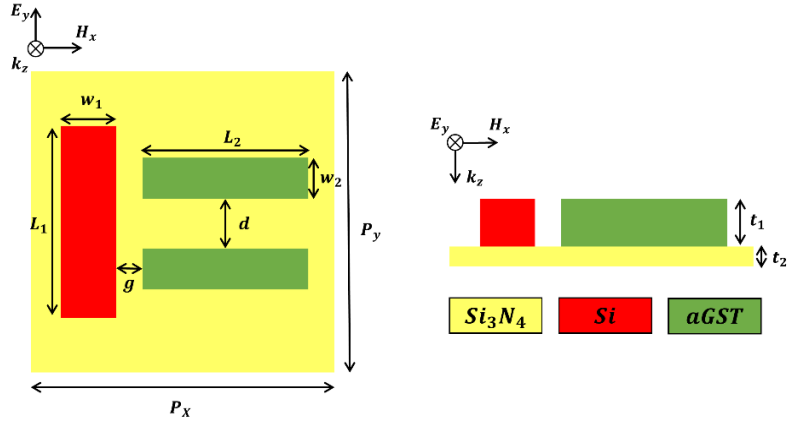

Fig.1 The top and front views for the periodic unit cell of proposed all-dielectric hybrid MM.

The top and front views for the periodic unit cell of proposed MM are shown in Fig. 1, including one vertical silicon rod (in red) and a pair of horizontal aGST rods (in green). The geometric parameters are as follows: $L_1$= 700 nm, $w_1$= 200 nm, $L_2$= 600 nm, $w_2$= 150 nm, $g$= 40 nm, $d$= 150 nm, $P_x$= $P_y$= 1100 nm, $t_1$= 160 nm, $t_2$= 50 nm. Here, we focus on SWIR band, the refractive indices as well as extinction coefficients (n+ik) of silicon, aGST and cGST are given in Fig. 2 [29,33]. The substrate is chosen as silicon nitride with n= 1.98 [33]. The effective dielectric constants of GST with different degrees of crystallinity are calculated following the Lorentz-Lorenz relation [30]:

$$\frac{\varepsilon_{GST}(\lambda,C)-1}{\varepsilon_{GST}(\lambda,C)-1} = C \times \frac{\varepsilon_{cGST}(\lambda)-1}{\varepsilon_{cGST}(\lambda)+2} + (1-C) \times \frac{\varepsilon_{aGST}(\lambda)-1}{\varepsilon_{aGST}(\lambda)+2} \qquad (1)$$

where $\varepsilon_{aGST}$ and $\varepsilon_{cGST}$ are dielectric constants of aGST and cGST, respectively. $C$ is the crystallinity of GST, ranging from 0% to 100%.

The finite-difference time-domain (FDTD) method is used to analysis the optical response of proposed structure. Periodic boundary conditions are set along $x$ and $y$ directions, and perfectly matched layers (PMLs) are adopted in the $z$ direction. To ensure the accuracy, the mesh size is set as 10 nm. One light source with $y$-direction polarization is placed on the top of the MM, and one monitor is put at the bottom of the substrate to detect the power and calculate the transmission spectra, one monitor is put upper the source to detect the power and calculate the reflection spectra

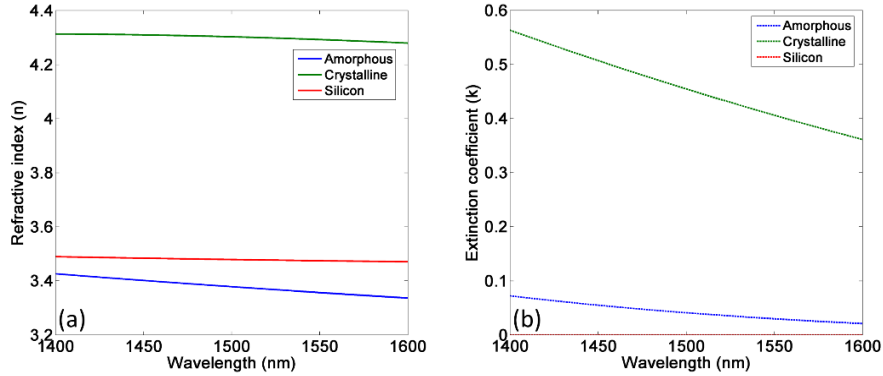

Fig. 2(a) The refractive indices n of aGST, cGST and silicon in SWIR band from 1400 and 1600 nm. (b) The extinction coefficients k of aGST, cGST and silicon in SWIR band.

## 3. Results and discussions

The transmission spectrum of proposed MM is given in Fig. 3(a), showing an EIT-like profile. It is known that such effect comes from a Fano-like resonance. The collective oscillations of the silicon rod, which can form an electric dipole mode, serve as bright mode. Such mode can couple strongly to normally incident light with *y*-direction polarization. The aGST rod pair can support an electric quadrupole dark mode, which cannot be stimulated directly by the free-space excitation with the electric field oriented along the *y* axis, but can couple to the bright mode through near-field coupling, as depicted in Fig. 3(b-c) [34]. The destructive interference between bright and dark mode with close resonant wavelengths can construct a typical three-level Fano-resonant system as indicated in the inset of Fig. 3(a), leading to the EIT-like spectrum with the transparent peak at 1500 nm [5].

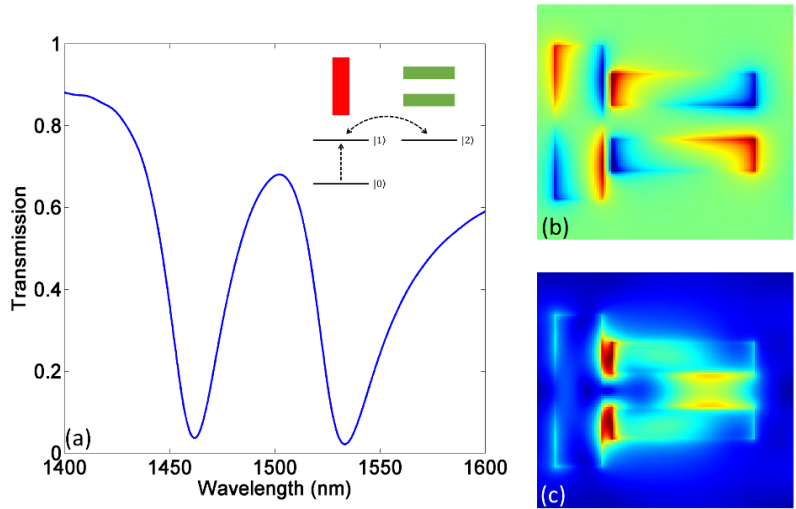

Fig. 3(a) The EIT-like transmission spectrum of proposed MM. The inset shows the typical three-level Fano-resonant system. (b) The distribution of z component of electric field $\mathbf{E}_z$ at 1500 nm, indicating the electric dipole mode in silicon rod and electric quadrupole mode in aGST rods. (c) The power distribution at 1500 nm, indicating clearly the near-field coupling between bright and dark mode.

Via the thermal, electrical or optical triggers, aGST can undergo the phase transition, turned into cGST. The transmission spectra under GST with different crystallinities are given in

Fig .4(a), showing that the interference-induced transparent peak has a red shift when the crystallinity increases. This is because the refractive index of GST will rise from amorphous to crystalline state, which lifts the resonant wavelength of the dark mode. Meantime, the peak will drop due to the increasing extinction coefficients of GST bringing about more intrinsic loss. Besides, the resonant wavelengths of bright and dark mode become more distinct, which also breaks the interference. Especially, the interference vanishes when GST comes to be 100% crystalline, leaving a resonant dip at 1500 nm, which indicating a pure bright mode. Fig 5(a-b) provide the power distribution when crystallinity is 50% and 100%. Obviously, combined with Fig. 3(c), the power in the gap will decline with the increase of the crystallinity, indicating a weaker coupling between the two modes at 1500 nm.

Fig. 4(b) provide transmission spectra under different gap distances *g* when GST is amorphous. The increasing gap distances will lead to a narrower and lower peak, which can attribute to the change of coupling strength between the two modes. We also investigate the corresponding transmission spectra when GST is fully crystalline, as presented in Fig. 4(c). Since this is no coupling between the two modes at 1500 nm in this case, the gap distances have little impact on transmission spectra.

Obviously, all-optical switch can be realized based on such mechanism. As shown in Fig. 4(d), the transmission at 1500 nm can be switched from 68% to 11% when aGST comes to cGST with gap distance 40 nm, and such phase transition can be stimulated optically by a pump light. The modulation depth (MD) is defined as follows:

$$MD = \frac{T_{max} - T_{min}}{T_{max}} \times 100\% \qquad (2)$$

Where $T_{max}$ and $T_{min}$ are maximum and minimum transmission at 1500 nm, respectively. The MD of such optical switch can reach 84%, showing a great modulation performance [35].

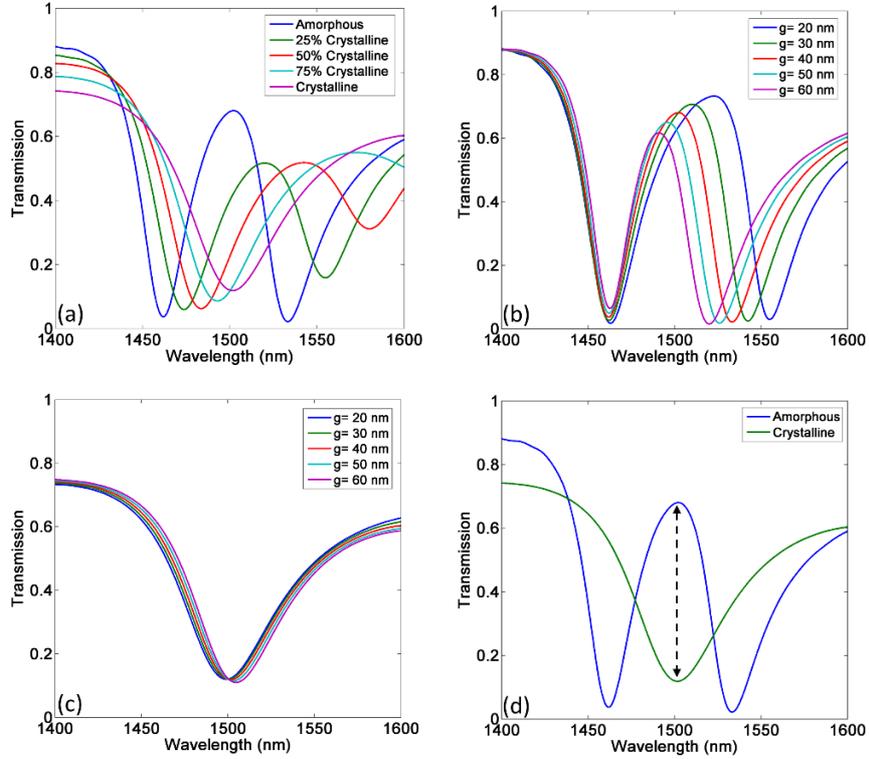

Fig. 4(a) The transmission spectra under GST with different crystallinities, *g*= 40 nm. (b) The transmission spectra under different gap distances when GST is amorphous. (c) The transmission spectra under different gap distances when GST is fully crystalline. (d) The transmission spectra under aGST and cGST with *g*= 40 nm.

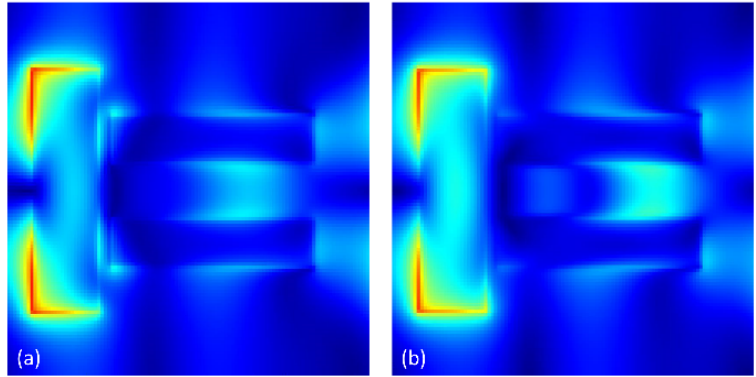

Fig. 5(a) The power distribution at 1500 nm when GST is 50% crystalline. (b) The power distribution at 1500 nm when GST is 100% crystalline.

Further, we propose that such MM can act as all-optical logic gates in the spatial domain, which can meet the requirement of next generation optical networks [36]. Here, we define input state, which is decided by the pump light, as *0* and *1* when GST is amorphous and crystalline respectively, as shown in Fig. 6(a). The transmission at 1500 nm is 68% and 11% respectively, corresponding to output state *1* and *0* as presented in Fig. 6(b). Consequently, the NOT gate is realized as given in table. 1. Moreover, if crystallinity of GST rod pair can be independently controlled, we can also introduce double input states as illustrated in Fig. 7(a). a/a GST, a/cGST, c/aGST and c/cGST rod pair corresponds to the input state *00*, *01*, *10* and *11*, respectively. Input state *00* leads to a high transmission 68% at 1500 nm which represents output state *1*, while the other input states lead to much lower transmission, which is output state *0*, as indicated in Fig. 7(b). Thereby, the NOR gate is efficiently established as shown in table. 2. At the same time, if we focus on the reflection spectra as shown in Fig. 7(c), the OR gate is also available as explained in table. 3. Thus, three different logic gates, NOT, NOR and OR, can be achieved in such device simultaneously.

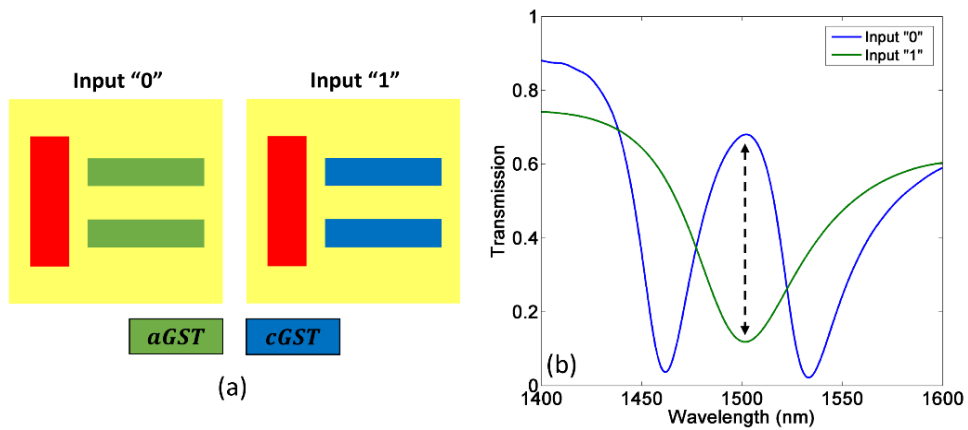

Fig. 6(a) The different GST states corresponding to different single input states. (b) The transmission spectra corresponding to different single input states.

**Table. 1 The NOT gate at 1500 nm.**

| Input states | Output transmission | Output states |
|---|---|---|
| *0* | 68% | *1* |
| *1* | 11% | *0* |

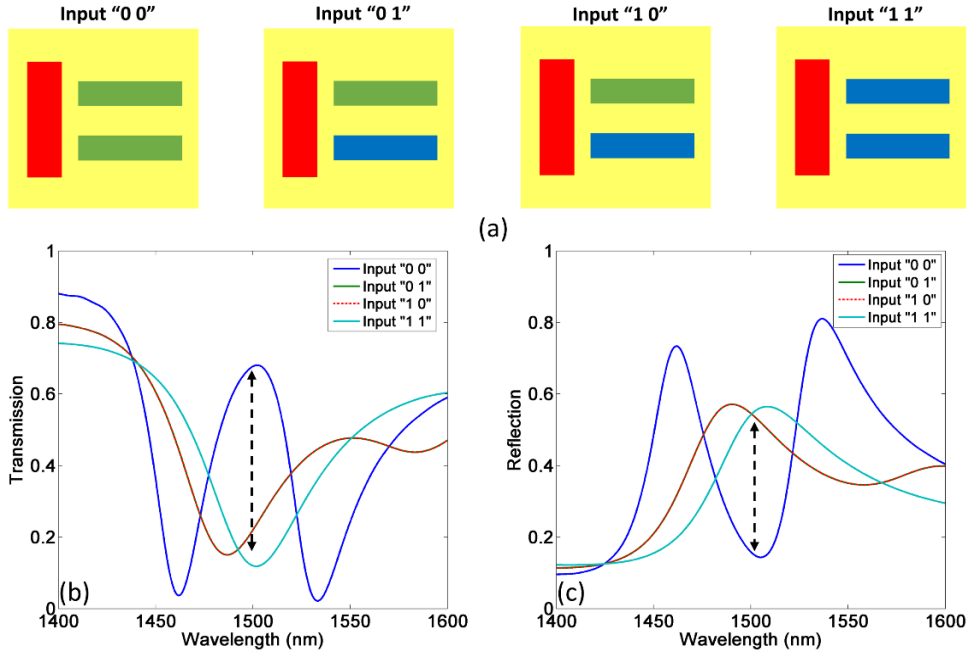

Fig. 7(a) The different GST states corresponding to different double input states. (b) The transmission spectra corresponding to different double input states. (c) The reflection spectra corresponding to different double input states.

**Table. 2 The NOR gate at 1500 nm.**

| Input states | Output transmission | Output states |
|---|---|---|
| *0 0* | 68% | *1* |
| *0 1* | 21% | *0* |
| *1 0* | 21% | *0* |
| *1 1* | 12% | *0* |

**Table. 3 The OR gate at 1500 nm.**

| Input states | Output reflection | Output states |
|---|---|---|
| *0 0* | 15% | *0* |
| *0 1* | 54% | *1* |
| *1 0* | 54% | *1* |
| *1 1* | 54% | *1* |

## 4. Conclusion

In summary, we proposed all-dielectric hybrid MM consisting of silicon and GST, and investigate its applications as all-optical switch and logic gates based on dynamically tunable

EIT, which benefits from the phase transition of GST. Considering such transition is fast, reversible and can be excited by light, the proposed device possesses reconfigurable capacity as well as ultrafast response time. Since the operation band is in SWIR, this device can find potential applications in next-generation optical information networks.

## Acknowledgments

This work is supported by the National Natural Science Foundation of China (61671455, 61805278), the Foundation of NUDT (ZK17-03-01), the Program for New Century Excellent Talents in University (NCET-12-0142), and the China Postdoctoral Science Foundation (2018M633704).